# Highly Efficient Carrier Multiplication in van der Waals layered Materials

Ji-Hee Kim[1,2†*], Matthew R. Bergren[1,2†], Jin Cheol Park[1,2†], Subash Adhikari[1,2], Michael Lorke[3], Thomas Fraunheim[4], Duk-Hyun Choe[5], Beom Kim[6], Hyunyong Choi[6], Tom Gregorkiewicz[7], and Young Hee Lee[1,2*]

[1]IBS Center for Integrated Nanostructure Physics (CINAP), Institute for Basic Science, Suwon 16419, Korea.

[2]Department of Energy Science, Department of Energy Science, Sungkyunkwan University, Suwon 16419, Korea.

[3]Institut fur Theoretishe Physik, Universitat Bremen, P. O. Box 330 440, 28334 Bremen, Germany.

[4]Bremen Center for Computational Materials Science, Institut fur Theoretishe Physik, Universitat Bremen, P. O. Box 330 440, 28334 Bremen, Germany.

[5]Department of Physics, Applied Physics and Astronomy, Rensselaer Polytechnic Institute, Troy, NY, USA.

[6]School of Electrical and Electronic Engineering, Yonsei University, Seoul 120-749, Korea.

[7]Van der Waals-Zeeman Institute, University of Amsterdam, Science Park 904, 1098 XH Amsterdam, The Netherlands.

[†]These authors contributed equally to this work.

[*]Correspondence to: kimj@skku.edu (J.-H.K), leeyoung@skku.edu (Y.H.L)

**Abstract:**

**Carrier multiplication (CM), a photo-physical process to generate multiple electron-hole pairs by exploiting excess energy of free carriers, is explored for efficient photovoltaic conversion of photons from the blue solar band, predominantly wasted as heat in standard solar cells. Current state-of-the-art approaches with nanomaterials have demonstrated improved CM but are not satisfactory due to high energy loss and inherent difficulties with carrier extraction. Here, we report ultra-efficient CM in van der Waals (vdW) layered materials that commences at the energy conservation limit and proceeds with nearly 100% conversion efficiency. A small threshold energy, as low as twice the bandgap, was achieved, marking an onset of quantum yield with enhanced carrier generation. Strong Coulomb interactions between electrons confined within**

**vdW layers allow rapid electron-electron scattering to prevail over electron-phonon scattering. Additionally, the presence of electron pockets spread over momentum space could also contribute to the high CM efficiency. Combining with high conductivity and optimal bandgap, these superior CM characteristics identify vdW materials for third-generation solar cell.**

**Main Text:** In the currently available solar cells, hot electron-hole pairs generated upon absorption of high-energy photons thermalize, thus converting their excess energy as heat and lowering the photovoltaic conversion efficiency. Harvesting this energy via efficient carrier multiplication (CM) would have a large impact by improving solar-cell power-conversion efficiency up to ~45%[1,2], above the Shockley-Queisser limit (~34%)[3]. In bulk semiconductors, CM or inverse Auger process can occur via impact excitation but this is an inefficient process with a high CM threshold energy of 4-5 times the bandgap ($E_g$) and low CM conversion efficiency[4,5]. The low efficiency of CM process in bulk originates from the density of final states, which is limited by the crystal momentum conservation, and also from the rapid carrier cooling which dominates over impact ionization in bulk materials.

Alternatively, low dimensional nanostructures relax strict momentum conservation and possess the discretized energy levels that limit carrier cooling[5-8]. Furthermore, strong Coulomb interaction in a layer-confined nanostructure can enhance inverse Auger recombination (Fig. S1). The CM conversion efficiency heavily depends on competition between the electron-electron (*el-el*) scattering rate and the electron-phonon (*el-ph*) cooling rate of hot carriers[9]. In nanostructures, electron-hole pairs exist as excitons and thus the CM is often referred to as the *multiple exciton generation*.

The efficiency and the threshold energy of a material are generally influenced by the following factors: i) Coulomb interactions, as promoted by spatial confinement and dielectric screening, ii) carrier cooling by phonon scattering, iii) initial/final density of states, and (possibly also) iv) surface passivation. While two-dimensional (2D) vdW layered materials have recently drawn much attention for quantum mechanical tunneling phenomena in ultrathin devices, they are also of particular interest from the CM point of view because of strong Coulomb interaction within spatially confined ultrathin (3-4Å) vdW layers, reduced dielectric screening that lead to an exciton binding energy of several hundred meV, much larger than those of quantum wells, and weak exciton-phonon coupling, which could lead to lower carrier cooling rates[10-20]. Therefore, these materials could be advantageous for promoting CM efficiency, and open a possibility to maximize the power conversion efficiency of solar cells to the bandgap range of 0.7-1 eV[2].

Here, we report CM phenomenon in transition metal dichalcogenide (TMdC) vdW layered materials. For the representative two materials, $MoTe_2$ and $WSe_2$, we demonstrate small CM threshold energy, as low as twice the bandgap, and high CM conversion efficiency of nearly 100%, markedly distinct from previous CM results in nanostructures and bulk[4,5,7-9,21]. Since $MoTe_2$ possesses a bandgap of 0.85 eV, ideal for CM in solar cells, the current findings propose this material as a strong candidate for maximum power conversion in solar cell. These results open a new route for vdW materials to be utilized for advanced light harvesting technologies, and in particular the 3rd generation photovoltaics of the future.

Transient absorption spectroscopy, monitoring the optically generated carrier dynamics arising from interband and intraband, has been successfully applied to study CM in semiconductor nanocrystals. Interband photo-bleaching (PB) via state-filling is a common technique to measure CM for direct bandgap materials[22,23]. For indirect bandgap materials, however, it is difficult to detect the photo-bleaching because of the weak interband absorption and lack of well-defined, resonant absorption features. Thus, photo-induced absorption (PIA) change caused by the transition within intrabands has been employed for the detection of CM in indirect bandgap materials[5,6]. In our study we perform PIA as well as PB measurements for comparison to measure the CM efficiency.

The transient absorption measurement we performed is schematically drawn in Figure 1a. A strong pump pulse excites carriers from the valence to the conduction band. A weak probe pulse, with a time delay, monitors the transmission change as a function of pump-probe delay time. A frequency-tunable broadband pulse was used for the pump pulse and the differential transmittance in the frequency domain was directly measured via probe pulse. Here, $\Delta T/T_0$ is defined as $(T_{on} - T_{off}) / T_{off}$, where $T_{on}$ and $T_{off}$ are the transmission of the probe with and without pump, respectively. The photo-induced absorption (PIA) and photo-bleaching (PB) are observed as the transmission change with respective negative and positive sign (Fig. 1b). Infrared probe energy is used in PIA, while the direct bandgap energy is used as a probe energy in PB. The rise components of PIA and PB in $\Delta T/T_0$ contain similar information of carrier population. PIA collects information of carrier population from multiple electron pockets, whereas PB collects information only at the electron pocket fixed at the probe energy. In this sense, PIA is more informative than PB to represent carrier populations in indirect bandgap materials (See SI, Fig. S2).

The steady-state absorption spectrum for a 16.4 nm thick 2H-MoTe$_2$ thin film sample grown by chemical vapor deposition[24] is shown in Figure 1c (see Fig. S3 and S4 for material characterization). Two main peaks, with an energy difference of 330 meV, were clearly resolved in the spectrum, corresponding to A and B exciton transitions (following Wilson and Yoffe's nomenclature) that arise from spin-orbit coupling. A direct excitonic bandgap near 1.05 ± 0.03 eV (marked as A in Fig. 1c) and an indirect bandgap of 0.85 ± 0.03 eV (marked as E$_g$ in Fig. 1c) were in good agreement with the calculated band structure in Fig. 1d, as well as reported values[25]. To study the CM properties of 2 × 2 cm$^2$ MoTe$_2$ thin films, we investigated the fluence-dependent differential transmittance spectra under different excitation energies.

Figure 2a describes low-fluence TA kinetics at an excitation energy below 2E$_g$ (1.58 eV or 780 nm). The hot electrons generated by pump pulse are detected by probe pulse via induced absorption as a function of delay time. The probe energy of 0.24 eV (5200 nm) is chosen because of better signal to noise ratio of differential transmittance in PIA spectrum. The choice of probe energy is rather indifferent to the determination of CM efficiency (See Fig. S5 and S6). Since differential transmittance varies with pump energy, $\Delta T/T_0$ is normalized by the absorbed photon density. The maximum $\Delta T/T_0$ increases with pump fluence, which corresponds to the density of photoexcited carriers mostly relaxed down to the electron pockets. When normalized by the absorbed photon density, they all fall into the same curve as shown in the inset, implying the absence of multiphoton absorption for the excitation energy of 1.58 eV (See also Fig. S7). The maximum intensity could be altered by the photocharging effect but in our case no such effect was observed (See Fig. S8).

It is obvious to see in Fig. 2b that the maximum intensity increases nearly twice that at excitation energies of 2E$_g$<E<3E$_g$ and three times at excitation energies of 3E$_g$<E<4E$_g$, compared to the maximum intensity at an excitation energy less than 2E$_g$. This reflects clearly a quantum yield of impact ionization. This is again strongly supported by the fact that the maximum intensity did not alter when excitation energies less than 2Eg are used, as demonstrated in the inset. Therefore, we can use the maximum differential transmittance as a measure of quantum yield of impact ionization.

We then followed typical procedure of determining the quantum yield[4-6,26]. In Fig. 2c, we plotted the maximum intensity as a function of fluence at different excitation energies. At energies lower than $2E_g$ (1.27, 1.58, 1.61 eV), they all fall into a single linear slope, which is regarded as a quantum yield of 1 and no carrier multiplication process occurred. Interestingly, the higher excitation energies, the steeper the linear slope. Thus, the quantum yield for carrier multiplication is represented by the ratio of the slope with respect to the linear slope at energy less than $2E_g$. The choice of maximum value of differential transmittance could be arbitrary and may influence the quantum efficiency. To check this, we chose the differential transmittance at a later delay time of 5 ps (See Fig. S12). Although the absolute slopes became smaller owing to the reduced intensity, the change in the ratio of the slope is not appreciable. This again confirms that information of carrier multiplication is included in a wide range of delay time.

The maximum peak is linearly proportional to the number of photo-generated carriers per absorbed photon that may involve CM when subjected to a higher pump energy larger than $2E_g$. The maximum induced absorption peak values ($\Delta T_{max}/T_0$) acquired at 320 fs after photoexcitation are plotted as a function of absorbed fluence for various photon energies (Fig. 2b), which increase linearly with the fluence. During the time to reach the maximum (rise time), the carrier-carrier scattering and carrier cooling processes occur and reach the bottom of the electron pockets. Each linear curve with the absorbed fluence ($F_{abs}$) for all excitation energies within the given fluence range is evidence for the absence of higher-order recombination on the timescale of our measurement such as many-body interactions (eg, exciton-exciton).

$\Delta T_{max}/T_0$ is related to the quantum yield via absorbed fluence ($F_{abs}$); $\Delta T_{max}/T_0 = \varphi\, \sigma_{intra}\, F_{abs}$, where the proportionality constants are the absorption cross section of the probe ($\sigma_{intra}$) and the quantum yield or carrier generation yield ($\varphi$)[4]. To determine the absorption cross section $\sigma_{intra}$, $\Delta T_{max}/T_0$ was measured at a low pump energy (1.58 eV, $1.85E_g$) with a given number of absorbed photons per unit area where CM is excluded ($E_{pump} < 2E_g$), and thus the quantum yield is presumably unity. This was averaged with several other energies (1.27 and 1.61 eV, $1.49E_g$ and $1.89E_g$). Solving the above equation, the absorption cross-section was determined to be $\sigma_{intra} = 1.24 \pm 0.02 \times 10^{-16}$ cm$^2$. Using this value, we then determined $\varphi$ by linear fitting of $\Delta T_{max}/T_0$ vs. $F_{abs}$ curve for each pump energy greater than $2E_g$. The maximum value of $\Delta T_{max}/T_0$ is proportional to the pump energy (Fig. 2b), yielding steeper slope for higher energy[5].

Figure 3a shows the extracted carrier generation quantum yield, normalized to the CM-free value, as a function of the pump energy, normalized by the 0.85 eV indirect bandgap energy of 2H-MoTe$_2$. From this plot, a direct determination of the CM efficiency ($\eta_{CM}$) of the material can be made, following the description proposed in Ref.[9,27]; $h\nu/E_g = 1 + \varphi/\eta_{CM}$, where $h\nu$ is the incident pump photon energy. A step-like feature with a threshold energy of ~$2E_g$ is clearly manifested. Quantum yield increases abruptly for pump photon energies exceeding ~$2E_g$. The obtained CM conversion efficiency $\eta_{CM}$ is ~99%. To investigate the phenomenon of ultra-efficient CM more generally for 2D vdW materials, a similar experiment was conducted for 2H-WSe$_2$ which has an indirect bandgap of 0.9 eV. The similar low threshold energy of ~$2E_g$ is observed, with the corresponding CM conversion efficiency of ~97%. The step-like behavior with a threshold energy of $2E_g$ seems to be a common feature for both materials and likely to be more general for 2D vdW dichalcogenides.

To confirm the validity of our finding obtained by the PIA method, we also conducted the PB measurements for 2H-MoTe$_2$ film sample, in a similar manner to PIA. The obtained quantum yield from PB is plotted in Fig 3a, which shows also a step-like behavior with a threshold energy of ~$2E_g$, and a CM conversion efficiency, ~93%. The quantum yield determined from PB is somewhat lower than that obtained from PIA, while the threshold energy is very similar, at $2E_g$. The lower quantum yield can be attributed to the fact that PIA monitors the total number of carriers, from multiple electron pockets, whereas PB collects the information of a specific transition only (at the direct bandgap). This implies that PIA is more informative than PB for the indirect bandgap materials. We also performed a control experiment of PIA for indirect bulk Si for comparison, with excitation energies at ~$1.4E_g$ and ~$2.8E_g$; no enhancement of carrier generation yield has been observed, in agreement with previous investigations of impact excitation in bulk Si which reported the excitation threshold energy as Exec > ~4eV. This confirms the validity of our experimental approach (See also Fig. S11).

Analysis of CM via maximum differential transmittance in PIA dynamics has advantages in studying 2D materials with multibands including indirect bandgap (Fig. S2) but the poor temporal resolution of infrared probe energy in our system limits further analysis with the rise time. We next studied the rise time of the bleach signal[22] and another experimental setup was used with a higher temporal resolution of 50 fs at 1.55 (< $2E_g$) and 3.1 eV (> $2E_g$) pump energies. The bleach, measured at the direct transition of A exciton (see Fig. 1c), rises quickly and is saturated for < $2E_g$ due to hot carrier relaxation through different bands or other processes (Fig. 3b). The delayed rise time is observed for the pump energy of > $2E_g$. We attribute this additional carrier generation time contributed from CM process, as shown in the schematics. The higher quantum yield is obtained at $3.65E_g$.

The subsequent decay features two components. We tentatively assign these decay components to carrier-carrier (faster, $\tau_1$) and carrier-phonon (slower, $\tau_2$) scattering. The $\tau_1$ decreases from 0.67 ps to 0.12 ps for > $2E_g$ incident pump energy, whereas the change of $\tau_2$ is not appreciable. This shortening of $\tau_1$ could appear due to the prevailing fast carrier-carrier Coulomb scattering. The experimentally determined fast decay time could be limited by the temporal resolution of the setup[25]. We note that the ratio of $\tau_2/\tau_1$ is similar to values reported previously for other 2D materials (See Table S1). Therefore, our assignment of $\tau_1$ to carrier-carrier scattering time, although some additional carrier-phonon scattering component in $\tau_1$ may be involved in our limit of temporal resolution, does not change our assignment (see also Fig. S9).

As mentioned before, the CM efficiency is governed in principle by the competition between carrier-carrier and carrier-phonon scattering. In general, the carrier-carrier scattering time constant in 2D materials is faster than materials in 0D (QDs) and 1D (see Table S1); the ratio of time constants ($\tau_2/\tau_1$) is about 30. Another primary factor for high CM conversion efficiency is that the carriers are bound between vdW layers. Even in multilayer vdW materials, carriers are confined within the potential barriers located between layers[28]. Therefore, we expect fast carrier-carrier scattering time. In other words, carrier-carrier scattering beats carrier-phonon scattering and consequently, phonon cooling is suppressed efficiently in 2D materials, yielding in highly efficient CM.

In discussion, we compare the QY results of various semiconductors in different geometries and compositions in Figure 4. The geometry of a material has a dramatic effect on the CM efficiency[4,6-8,29-31]. For example, PbS nanoplatelets and QDs are more efficient than their bulk counter-part[4,8]. Material composition also has a large effect on both threshold energy and CM efficiency. A step-like feature with threshold energy of $2.5E_g$ and CM efficiency of 90% is observed from Si QDs[6], while no such a feature is observed in PbS QDs. The most efficient CM materials measured to date have been quantum dots with a threshold energy of over $2.5E_g$ or $3E_g$ but with the CM efficiency below 90%. Recent reports for PbSe QD films[32] and improved CM efficiency in nanorods[8,21], and InAs QDs[29] show a threshold in the vicinity of $2E_g$ but poor CM efficiency of 35-80%. In contrast, the vdW layered materials investigated here clearly manifest a threshold energy of $2E_g$ with no excessive energy and the record CM conversion efficiency of nearly 100%. This is ascribed to the strong confinement effect by the electrostatic potential built within extremely narrow region of 3-4 Å[28]. We emphasize that our method of extracting the quantum yield via the comparison of maximum differential transmittance in PIA or in PB is different from that via the comparison of carrier population at short and long delay time in PB, frequently used in past research on CM in quantum confined systems. In our approach PIA and PB methods provide different quantum yield for indirect bandgap materials since the former extracts information of carrier population at both direct and indirect bands while the latter extracts information only at direct band. Meanwhile, both PIA and PB extract the similar information for direct band materials.

In summary, we demonstrate highly efficient CM in 2H-MoTe$_2$, with a lowest threshold energy and an ideal CM conversion efficiency. The photon energy in excess of the bandgap is not wasted by heat, and rapid electron-electron scattering prevails over electron-phonon scattering due to strong Coulomb scattering. Additionally, the presence of electron pockets in momentum space could also contribute to the high efficiency. Combining with high conductivity, large absorbance and optimal bandgap, these superior CM characteristics identify 2D vdW materials including MoTe$_2$ as a most promising new material for third generation solar cells, although many hurdles will have to be overcome for practical applications.

**Acknowledgements** This work was supported by the Institute for Basic Science (IBS-R011-D1).

**Author Contributions** Y.H.L. conceived the research. J.-H.K. and M.R.B. designed the experiment and performed the time-resolved experiments including data analysis and simulation. J.C.P grew and prepared the sample by wet transfer. J.C.P. and S.A. characterized the samples. M.L., T.F. and D.-H.C. carried out theoretical calculations. J.-H.K, B.K. and H.C.

performed optical pump-probe spectroscopy with the higher temporal resolution. J.-H.K., M.R.B, T.G. and Y.H.L interpreted the results, and wrote the manuscript. All authors discussed the results and commented on the manuscript.

**Additional information** Supplementary Information is available in the online version of the paper. Reprints and permissions information is available online at [www.](www. ) . Correspondence and requests for materials should be addressed to J.-H.K. and Y.H.L.

**Competing financial interests** The authors declare that they have no competing interests.

high room temperature mobility. *Nano Res.* **7**, 1731 (2014).

11. Beal, A. & Liang, W. Excitons in 2H-WSe$_2$ and 3R-WS$_2$. *J. Phys. C: Solid State Physics* **9**, 2459 (1976).

12. del Corro, E. *et al.* Atypical Exciton–Phonon Interactions in WS$_2$ and WSe$_2$ Monolayers Revealed by Resonance Raman Spectroscopy. *Nano Lett.* **16**, 2363 (2016).

13. Koirala, S., Mouri, S., Miyauchi, Y. & Matsuda, K. Homogeneous linewidth broadening and exciton dephasing mechanism in MoTe$_2$. *Phys. Rev. B* **93**, 075411 (2016).

14. Moody, G. *et al.* Intrinsic homogeneous linewidth and broadening mechanisms of excitons in monolayer transition metal dichalcogenides. *Nat. Commun.* **6**, 8315 (2015).

15. Komsa, H.-P. & Krasheninnikov, A. V. Effects of confinement and environment on the electronic structure and exciton binding energy of MoS$_2$ from first principles. *Phys. Rev. B* **86**, 241201 (2012).

16. Chernikov, A. *et al.* Exciton binding energy and nonhydrogenic Rydberg series in monolayer WS$_2$. *Phys. Rev. Lett.* **113**, 076802 (2014).

17. He, K. *et al.* Tightly bound excitons in monolayer WSe$_2$. *Phys. Rev. Lett.* **113**, 026803 (2014).

18. Mitioglu, A. *et al.* Optical investigation of monolayer and bulk tungsten diselenide (WSe$_2$) in high magnetic fields. *Nano Lett.* **15**, 4387 (2015).

19. Yang, J. *et al.* Robust excitons and trions in monolayer MoTe$_2$. *ACS Nano* **9**, 6603 (2015).

20. Tielrooij, K. J. *et al.* Photoexcitation cascade and multiple hot-carrier generation in graphene. *Nat. Phys.* **9**, 248 (2013).

21. Cunningham, P. D. *et al.* Enhanced Multiple Exciton Generation in Quasi-One-Dimensional Semiconductors. *Nano Lett.* **11**, 3476 (2011).

22. Ellingson, R. J. *et al.* Highly Efficient Multiple Exciton Generation in Colloidal PbSe and PbS Quantum Dots. *Nano Lett.* **5**, 865 (2005).

23. Schaller, R. D., Agranovich, V. M. & Klimov, V. I. High-efficiency carrier multiplication through direct photogeneration of multi-excitons via virtual single-exciton states. *Nat. Phys.* **1**, 189 (2005).

24. Park, J. C. *et al.* Phase-Engineered Synthesis of Centimeter-Scale 1T'- and 2H-Molybdenum Ditelluride Thin Films. *ACS Nano* **9**, 6548 (2015).

25. Nie, Z. *et al.* Ultrafast Carreir Thermalization and Cooling Dynamics in Few-Layer MoS$_2$. *ACS Nano* **8**, 10931 (2014).

26. Trinh, M. T. *et al.* In Spite of Recent Doubts Carrier Multiplication Does Occur in PbSe
10

**Figure Caption**

**Figure 1. Pump-probe spectroscopy. a,** Schematic of the differential transmittance experiment. **b,** Conceptual absorption spectra with and without pump (top) and differential transmittance spectrum (bottom). Flat band model representing photobleaching and photoinduced absorption transition for photoexcited carrier (right). **c,** Absorption spectrum with various exciton peaks including primary A and B excitons in 2H-MoTe$_2$ thin film. An indirect bandgap ($E_g$) is also shown. The inset clearly reveals the exciton peaks after subtraction of the back ground absorption. **d,** Band structure and density of states for thin-film 2H-MoTe$_2$. **e.** Raw spectra of differential transmittance at different energy ranges (details shown in supplementary information S5).

**Figure 2. Carrier kinetics and variation of maximum differential transmittance with photon energies in 2H-MoTe$_2$ film. a,** Time-dependent photo-induced absorption signals at different pump fluencies with a pump energy of 1.58 eV. The kinetics with normalized intensity is shown in the inset, implying the absence of multiphoton absorption. **b,** PIA kinetics at different excitation energies (1.58, 2.38, and 2.76 eV), normalized to the equal number of absorbed photons. The inset show the normalized PIA kinetics excited below 2$E_g$ (1.59 eV, and 1.61 eV). **c,** Maximum $\Delta T_{max}/T_0$ as a function of absorbed fluence at different pump photon energies. The linear slope indicates quantum yield. The steeper the slope of the line, the higher the quantum yield.

**Figure 3. Quantum yield plot for 2H-MoTe$_2$ and WSe$_2$ films. a,** The QY data for 2H-MoTe$_2$ (red dots) and 2H-WSe$_2$ (hollow triangles) film are plotted as a function of excitation energy normalized by the bandgap of each material. The black solid line and the dashed lines in gray color represent simulations of CM efficiency ($\eta_{CM}$). The schematic demonstrates the photogenerated carriers with (top) and without (bottom) CM. The processes (1) and (2) represent hot carrier cooling via phonon emission and carrier multiplication, respectively. To compare PIA data with PB data, quantum yield obtained from PB for 2H-MoTe$_2$ film is added (blue diamonds). **b,** Rise and decay dynamics for two different excitation energies for a 2H-MoTe$_2$ film. The bleaching signal, taken at direct transition A, sharply rises for pumping at 1.82$E_g$, while the slower rise time is shown at 3.65$E_g$. The decay curves with the long time delay up to 1 ns are displayed in the inset.

**Figure 4. Quantum yield for various nanostructures and bulk materials and the extracted carrier multiplication.** Comparison of the CM efficiency of thin-film 2H-MoTe$_2$

and 2H-WSe$_2$ with various other semiconductors including bulk, QDs, and nanoplatelets. Quantum yields from previous reports were taken from Refs [4,6-8,29-31].

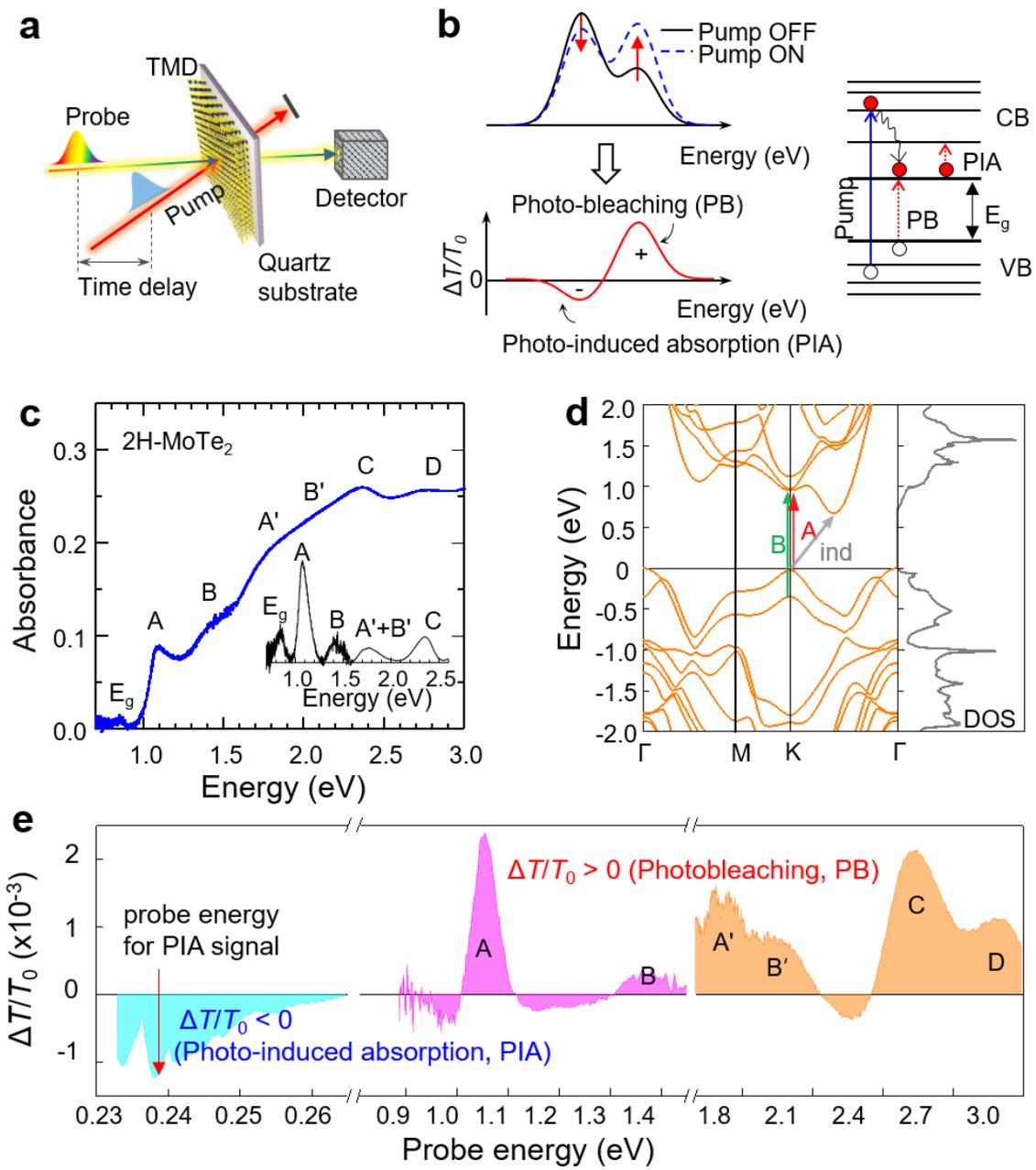

**Kim et al. Figure 1**

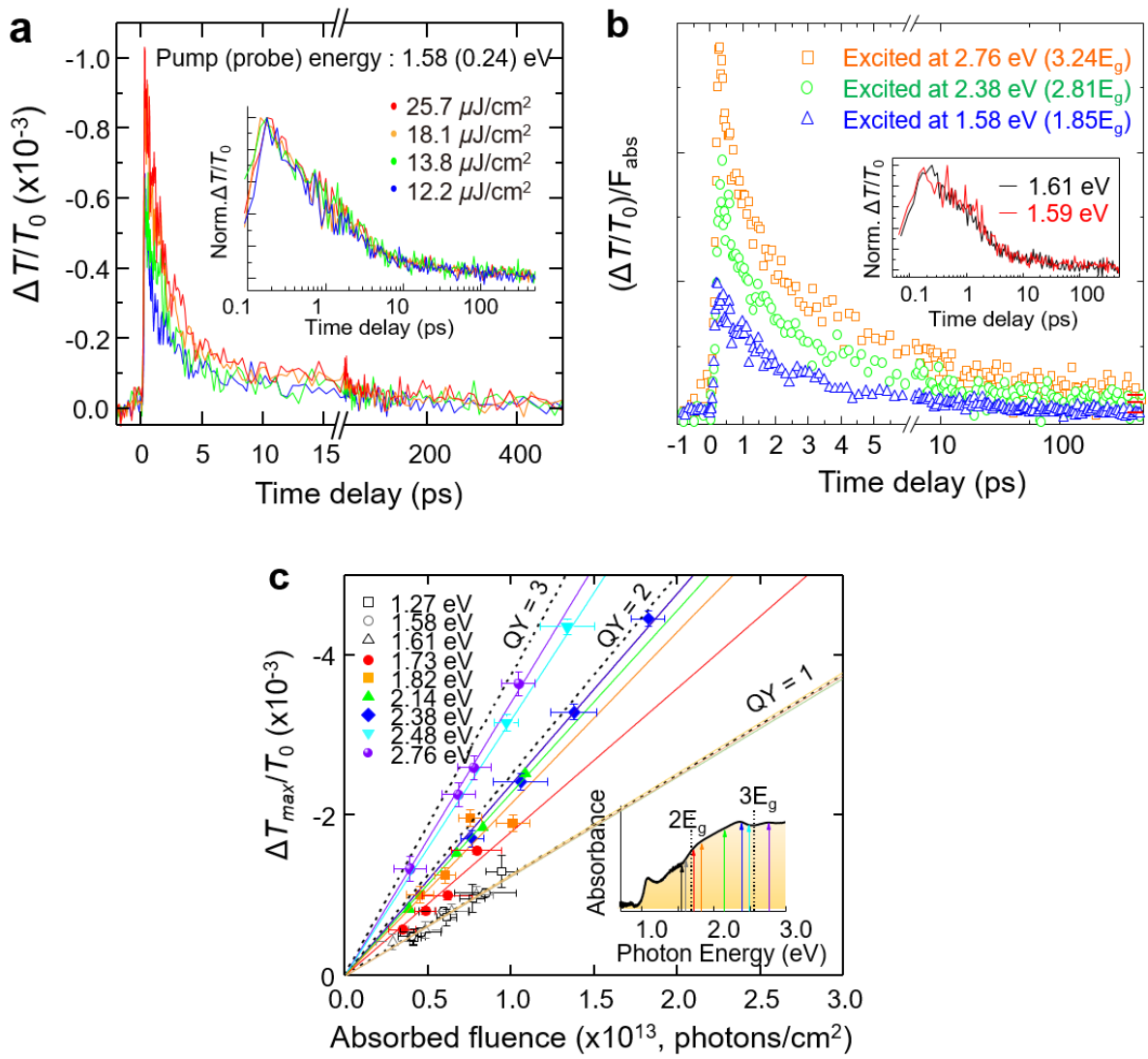

**Kim et al. Figure 2**

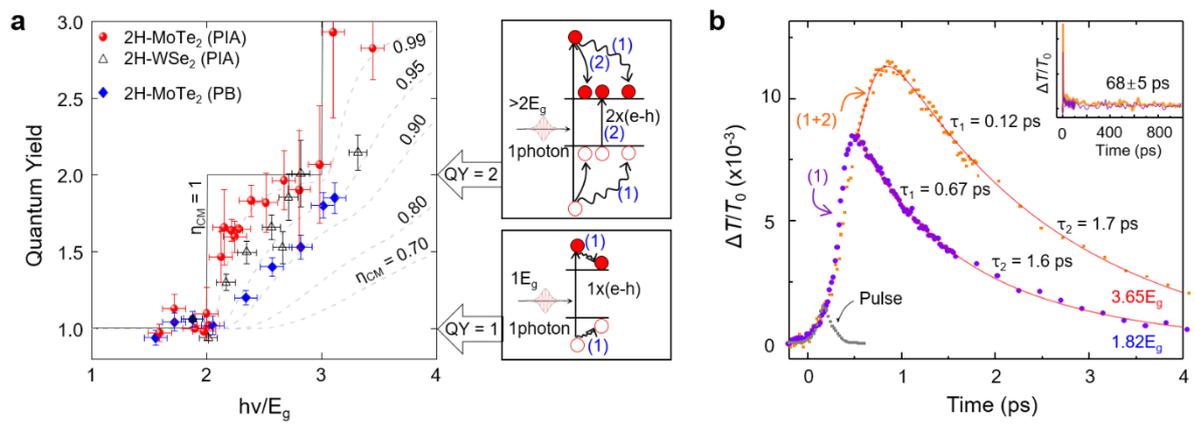

**Kim et al. Figure 3**

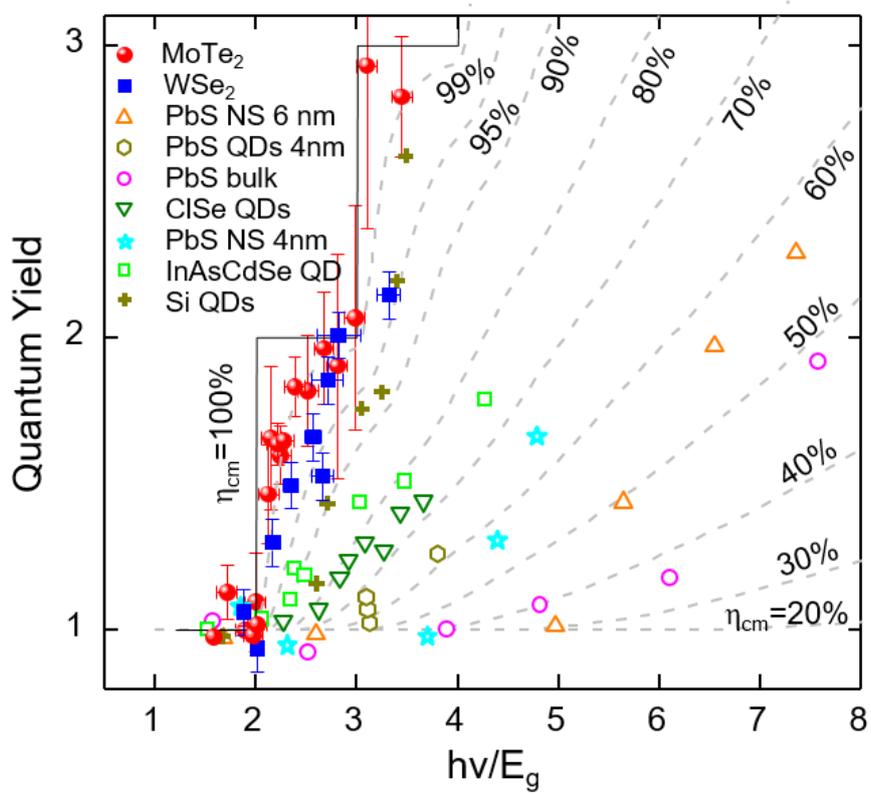

**Kim et al. Figure 4**